\documentstyle[11pt]{article}

\begin{document}

\title{Generalized uncertainty principles, effective Newton constant and regular black holes}
\author{Xiang Li\thanks{xiang.lee@163.com}\\
 {\it Department of Physics}\\
 {\it Jimei University, 361021, Xiamen, Fujian province, P. R. China}\\
 Yi Ling\thanks{lingy@ihep.ac.cn}\\
 {\it Institute of High Energy Physics}\\
 {\it Chinese Academy of Sciences,
 100049,
 Beijing, P. R. China}\\
 You-Gen Shen\thanks{ygshen@shao.ac.cn}\\
 {\it Shanghai Astronomical Observatory}\\
 {\it Chinese Academy of Sciences,
 200030, Shanghai, P. R. China}\\
 Cheng-Zhou Liu\thanks{czlbj20@163.com}\\
{\it Department of Physics}\\
{\it Shaoxing University, 312000, Shaoxing, Zhejiang province, P. R.
China}\\
Hong-Sheng He\thanks{hsh@jmu.edu.cn}\\
{\it Department of Physics}\\
{\it Jimei University, 361021, Xiamen, Fujian province, P. R.
China}\\
Lan-Fang Xu\thanks{lance\_xu01@163.com}\\
{\it Department of Physics}\\
{\it Jimei University, 361021, Xiamen, Fujian province, P. R.
China}\\}

\date{}
\maketitle
\newpage

\begin{abstract}
In this paper, we explore the quantum spacetimes that are
potentially connected with the generalized uncertainty principles.
By analyzing the gravity-induced quantum interference pattern and
the Gedanken for weighting photon, we find that the generalized
uncertainty principles inspire the effective Newton constant as same
as our previous proposal. A characteristic momentum associated with
the tidal effect is suggested, which incorporates the quantum effect
with the geometric nature of gravity. When the simplest generalized
uncertainty principle is considered,  the minimal model of the
regular black holes is reproduced by the effective Newton constant.
The black hole's tunneling probability, accurate to the second order
correction, is carefully analyzed. We find that the tunneling
probability is regularized by the size of the black hole remnant.
Moreover, the black hole remnant is the final state of a tunneling
process that the probability is minimized. A theory of modified
gravity is suggested, by substituting the effective Newton constant
into the Hilbert-Einstein action.

\vspace{1cm} {\bf Keywords}: generalized uncertainty principle,
effective Newton constant, characteristic momentum, regular black
hole, quantum tunneling, WKB approximation.
\end{abstract}
\newpage
\section{Introduction}

On the ground of dimensional analysis\cite{mtw}, the Planck length
is defined as $\ell_\textsc{p}\equiv\sqrt{\hbar G/c^3}$, where $c$
is the speed of light, $G$  Newton constant, and $\hbar$ Planck
constant. This unit of length should appear in  any theory
reconciling general relativity and quantum theory. It is generally
believed that $\ell_\textsc{p}$ is the shortest measurable length,
and quantum gravity effects( or quantum fluctuations in spacetimes)
become crucial to understanding the physics on this length scale. As
a classical theory, general relativity involves only $c$ and $G$,
and the minimal length cannot be predicted naturally by the theory
itself. There are some problems are somewhat related to the defect
that the Planck length is absent in the classical spacetimes.

 One  of problems is the
spacetime singularity\cite{ph}. Following from Penrose and Hawking's
theorems, the spacetime singularity is inevitable in the framework
of classical general relativity. In a certain extent, the
singularity is characterized by the divergence of Kretschmann
scalars ($K^2=R_{\rho\lambda\mu\nu}R^{\rho\lambda\mu\nu}$). For a
Schwarzschild black hole, $K^2\sim M^2/r^6$ become divergent, as
$r\rightarrow 0$.

Another problem is the fate of black hole evaporation\cite{hawk}. In
the case of Hawking's temperature expression( $T_H\sim M^{-1}$), the
negative capacity( $C\sim -M^{-2}$) makes the black hole evaporation
faster and faster. Both the temperature and the mass loss rate(
$\dot{M}\sim -M^{-2}$) become divergent, if the black hole vanishes.

The third problem is related to the tunneling picture of the black
hole radiation\cite{parik,parikh2,vagenas4,vagenas1, zjy2,zjy1}. The
tunneling probability accurate to the first order correction becomes
explosive\cite{vagenas1,zjy1}, if the final size of the black hole
is allowed to be arbitrary small. When the second order correction
to the tunneling probability is considered, the situation becomes
worse\cite{zjy1}. We are confronted with an unacceptable picture
that a black hole of any mass could vanish in an instant. This
difficulty is associated with the absence of the minimal length, and
then it is
 not necessarily overcome by improving the WKB method.

These problems are expected to be solvable at the presence of
quantum gravity effects. The generalized uncertainty
principle(GUP\cite{garay1}), as one of methods of quantum gravity
phenomenology, has been applied to the black hole thermodynamics in
some heuristic manners\cite{pschen1,vagenas2, lx1}. GUP imposes a
lower bound on the size of the black hole, and modifies the black
hole's thermodynamics. However, it is hard to understand that the
temperature approaches a value of order of Planck temperature, while
the heat capacity of the minimal black hole vanishes. In other
words, GUP predicts the existence of black hole remnant, but the
temperature puzzle has not been solved completely.

 This dilemma may be associated with such a working
hypothesis in the literature that the matter is dominated by GUP,
while the spacetimes  are classical.  This hypothesis can be named
as the GUP-revised semiclassical theory. However, the corrections to
classical spacetimes should be considered seriously, especially on
the Planck scale. After all, a spacetime dominated by quantum
gravity effects may be essentially different from the classical and
smooth background. The quantum spacetime should reflect the
existence of the minimal length, when the GUP is considered in an
appropriate manner.

 Some regular black holes with finite Kretschmann
scalars have been suggested in the
literature\cite{bardeen,reuter,nicolini,hayward,myung,lx2,culetu,vagenas3,ghosh},
and they give rise to the zero temperature remnants. As a tentative
attempt, we suggest a regular black hole which is connected with the
GUP by an effective Newton constant\cite{lx2}. This suggestion is
based on an observation upon the role that the GUP plays in the
relation between the gravitational acceleration and Newton
potential, in the context of operators. The effective Newton
constant is motivated by substituting the GUP for Heisenberg
commutator. However, this direct substitution of commutators is in
the shortage of a clear physical picture. Moreover, it not
consistent with the GUP-modified Hamiton equation\cite{chang1}, and
its reliability should be checked by other methods. We expect that
the effective Newton constant can be inspired by the GUP in some
concrete physical processes. Another limitation of our previous work
is that the simplest GUP(as presented in the next section) doesn't
give rise to a regular black hole, although it means the existence
of the minimal length. This shortage is related to such a
characteristic momentum as  $\Delta p\sim\hbar/r$. Although this
momentum scale is motivated by Heisenberg's principle, it  is
irrelevant to the quantum fluctuations  in spacetime,  because it
doesn't involve the  Newton constant and the mass of source of
gravitational field.

The aim of this paper is to gain a better understanding of the
GUP-inspired effective Newton constant and quantum spacetimes, at
the level of quantum gravity phenomenology. Firstly, the effective
Newton constant inspired by the GUP will be reexamined in two
Gedankens involving gravitation and quantum theory, i.e. COW phase
shift and Einstein-Bohr's box. Secondly, the momentum scale will be
reconsidered. In our opinion, an appropriate scale  should reflect
the amplitudes of the quantum fluctuations in the curved spacetime,
and may be related to the geometric character of gravity.  It would
be different from that suggested
 in Ref.\cite{lx2}. Thirdly, we will consider a regular black hole,
which is inspired by the simplest GUP, i.e. the most popular
version. Finally, quantum radiation from this black hole will be
discussed seriously in this work. For a mini black hole, the
temperature may lose its usual meaning in the thermodynamics. It is
more reasonable to consider the quantum tunneling from this black
hole. We are interested in the role that the minimal length plays in
the tunneling process, and in the question of whether the explosion
of tunneling probability occurs in the quantum spacetime.

\section{Effective Newton constant inspired by the generalized uncertainty principles}
Based on some theoretical considerations and gedanken experiments
for incorporating gravitation with quantum theory,
 Heisenberg's uncertainty principle is likely to suffer a
 modification as follows\cite{garay1}
\begin{eqnarray}\label{gup0}
\Delta x\sim\frac{\hbar}{\Delta
p}+\frac{\alpha\ell_\textsc{p}^2}{\hbar}\Delta p,
\end{eqnarray}
Where $\alpha$ is a dimensionless number of order of unity. Since
GUP means the minimal length of order of $\ell_\textsc{p}$, it
should be crucial for the Planck scale physics. Corresponding to
(\ref{gup0}), Heisenberg's commutator is extended
to\cite{kempf,ahlu}
\begin{eqnarray}\label{commut0}
[\hat{x},\hat{p}]=i\hbar\left(1+\frac{\alpha\ell_\textsc{p}^2}{\hbar^2}\hat{p}^2\right),
\end{eqnarray}
which will be considered seriously in this paper. However, there are
some other types of the generalized uncertainty
principles\cite{rama}\cite{hh}, and the commutation relations are
not necessarily the same as (\ref{commut0}). So we begin with a more
general commutator as follows\cite{lx2}
\begin{eqnarray}\label{commut1}
[\hat{x},\hat{p}]=i\hbar z,
\end{eqnarray}
where $z=z(\hat{p})$ is a function of momentum. The average value of
$z$, should ensure a lower bound on the measurable distance,
\begin{eqnarray}\label{uncertainty2}
\Delta x\geq\frac{z\hbar}{\Delta p}\geq\ell_\textsc{p},
\end{eqnarray}
which suggests the discreteness of the spacetime\cite{vagenas5}.
Considering the relation between Heisenberg's principle and the
wave-particle duality, let us derive the modified de-Broglie formula
from the generalized commutator (\ref{commut1}).
\subsection{Modified wave-particle duality}
It is well known that the proposal of uncertainty principle is
closely related to the wave particle duality.  Uncertainty relation
can be derived from de Broglie formula, by analyzing the
Heisenberg's microscope gedanken experiment or
 the single slit diffraction of light. However, once the framework of quantum mechanics is established
and Heisenberg's uncertainty relation is regarded as a fundamental
principle,  de Broglie formula becomes a deduction\cite{cdl}.
Concretely speaking, the momentum eigenstate
$\psi_p=\exp(ipx/\hbar)$ can be derived from the canonical
commutation relation $[\hat{x},\hat{p}]=i\hbar$, and de Broglie
formula is obtained by comparing the momentum eigenstate with a
plane wave function $\exp(2\pi ix/\lambda)$. It is expectable that
de Broglie formula should suffer a modification, when the
Heisenberg's commutator is changed. Corresponding to the generalized
commutator (\ref{commut0}), the modified de Broglie relation is
given by\cite{kempf,ahlu}
\begin{eqnarray}\label{mdb1}
\lambda=\frac{2\pi
\ell_\textsc{p}\sqrt{\alpha}}{\arctan(\ell_\textsc{p}\sqrt{\alpha}p/\hbar)}.
\end{eqnarray}
It is easy to check that Eq.(\ref{mdb1}) satisfies the following
relation
\begin{eqnarray}\label{mdbb}
\frac{d}{dp}\left(\frac{2\pi}{\lambda}\right)=\hbar^{-1}\left(1+\frac{\alpha\ell_\textsc{p}^2}{\hbar^2}{p}^2\right)^{-1}.
\end{eqnarray}
As argued in the following, a more general formula associated with
the commutator (\ref{commut1}) is given by
\begin{eqnarray}\label{mdb2}
\frac{d}{dp}\left(\frac{2\pi}{\lambda}\right)=\hbar^{-1}z^{-1}.
\end{eqnarray}
In order to explain this formula, we first construct a commutator as
follows
\begin{eqnarray}\label{commut2}
[\hat{x}, \hat{k}]=i,
\end{eqnarray}
where $\hat{k}=k(\hat{p})$ is a function of momentum operator.
Following from the law of operator algebra, we obtain
\begin{eqnarray}\label{commut3}
[\hat{x},\hat{p}]=[\hat{x},k]\frac{d\hat{p}}{dk}=i\frac{d\hat{p}}{dk}.
\end{eqnarray}
Comparing (\ref{commut1}) with (\ref{commut3}), we have
\begin{eqnarray}
\frac{dk}{d\hat{p}}=\hbar^{-1}z^{-1},\nonumber
\end{eqnarray}
and then
\begin{eqnarray}
k(\hat{p})=\hbar^{-1}\int z^{-1}d\hat{p}.
\end{eqnarray}
Obviously, $[\hat{p},k(\hat{p})]=0$, this means that there is a
common eigenstate $\psi_p$ of eigenvalue $p$, which satisfies
\begin{eqnarray}\label{kstate}
\hat{p}\psi_p&=&p\psi_p,\nonumber\\
k(\hat{p})\psi_p&=&k(p)\psi_p,
\end{eqnarray}
where
\begin{eqnarray}\label{mdb3}
k(p)=\hbar^{-1}\int z^{-1}(p)dp,
\end{eqnarray}
is the eigenvalue of the operator $k(\hat{p})$. Since
$[\hat{x},\hat{p}]\neq i\hbar$, the momentum operator is no longer
represented by $\hat{p}=-i\hbar\nabla$. However, comparing
(\ref{commut2})
 with Heisenberg's commutator, we obtain $\hat{k}=-i\nabla$.  For one dimensional case, the second
equation of (\ref{kstate}) becomes
\begin{eqnarray}
-i\frac{d\psi_p}{dx}=k(p)\psi_p.
\end{eqnarray}
So the momentum eigenstate is given by
\begin{eqnarray}
\psi_p=\exp(ikx),
\end{eqnarray}
which describes a plane wave of wavelength $\lambda=2\pi/k$. Thus
$k(\hat{p})$ introduced in (\ref{commut2})
 can be viewed as the wave-vector operator, and (\ref{mdb3}) is just the
 wave-number. The modified wave-particle
 duality is
 characterized by  (\ref{mdb2}) or (\ref{mdb3}), which is the basis for the following discussions.

\subsection{COW phase shift}
In 1975, Colella, Overhauser, and Werner(COW) observed the
gravity-induced quantum interference pattern of two neutron
beams\cite{cow1}. When the plane of two beams is vertical to the
horizontal plane, the phase shift is given by\cite{cow2,cow3}
\begin{eqnarray}\label{phase1}
\Delta\varphi=\frac{mgA}{\hbar v},
\end{eqnarray}
where  $g$ denotes the earth's gravitational acceleration,  $v$ the
average  speed of neutrons, and $A$  the area enclosed by two
interfering neutron beams that propagate on two paths on a plane.

This famous experiment may be regarded as a test of the property of
gravity in the microscopic world\cite{saha}. It is naturally
expected to shed light on  the quantum structure of spacetime, by
attaching the GUP's significance to the
 gravity-induced phase shift. In the following discussions, COW experiment will be revisited in a heuristic manner\cite{cow3}. Let us
 consider two interfering neutron beams. For simplicity, the plane
 of two beams is set to be vertical to the horizontal plane. The
 first (upper) beam propagates on a horizontal path and a vertical
 downward path in sequence. The
 second (lower) beam propagates on a vertical
 downward path and a horizontal path in sequence. The area enclosed by two beams is $A=yl$, where $l$ is the length of each horizontal
 path,  and $y$ is the height of the upper horizontal path with respect to the lower horizontal
 one. It is shown by simple analysis that the change in the phase of
 one vertical beam cancel out that of another vertical beam, and  then the phase shift is
 completely  attributed to the gravity-induced difference in the wavelength
 of
 two horizontal beams,
\begin{eqnarray}\label{phase2}
\Delta\varphi^{\prime}&=&2\pi \left(\frac{l}{\lambda_2}-\frac{l}{\lambda_1}\right)\nonumber\\
&=&l(k_2-k_1)\nonumber\\
&=&l\Delta k=l\frac{\Delta k}{\Delta p}\Delta p,
\end{eqnarray}
where $\Delta p=p_2-p_1$ is the difference in momentum of two
horizontal beams. Since  $\Delta p$ is a  small quantity,
Eq.(\ref{phase2}) can be expressed as
\begin{eqnarray}\label{phase3}
\Delta\varphi^{\prime}&\approx&l\frac{dk}{dp}\Delta p\nonumber\\
&=&\hbar^{-1}z^{-1}l\Delta p,
\end{eqnarray}
 where Eq.(\ref{mdb2}) has been considered. The neutron beams
 propagate in the earth's gravitational field, and obey energy
 conservation law, so we have
\begin{eqnarray}\label{energy1}
mgy&=&\frac{p_2^2-p_1^2}{2m}\nonumber\\
&=&v\Delta p,
\end{eqnarray}
where the earth's rotation is neglected,  and $v=(p_1+p_2)/2m$.
Substituting (\ref{energy1}) into (\ref{phase3}), we obtain
\begin{eqnarray}\label{phase4}
\Delta\varphi^{\prime}=\frac{mgyl}{z\hbar
v}=\frac{mg^{\prime}A}{\hbar v},
\end{eqnarray}
 where $g^{\prime}=g/z, A=yl$. When $z=1$, Eq.(\ref{phase4}) returns to (\ref{phase1}), which is  just the earlier result predicted by usual quantum theory.

As shown by Eqs.(\ref{phase1}) and (\ref{phase4}), the expression
for the corrected phase shift is almost the same as the usual
result, except a momentum-dependent factor $z$.  The latter can be
obtained from the former by replacing  $g$ with $g/z$.   The GUP's
significance to COW experiment is equivalent to the situation that
two neutron beams propagate in a modified gravitational field
characterized by the effective field strength $g^{\prime}=g/z$.

\subsection{Weighting photon}
In 1930, Einstein devised a subtle gedanken experiment for weighting
photon\cite{photon-box1, photon-box2}, and tried to demonstrate the
inconsistency of quantum mechanics. Einstein considered a box that
contains photon gas and hangs from a spring scale.  An ideal clock
mechanism in the box can open a shutter. Einstein assumed that the
ideal clock could determine the emission time exactly (i.e. $\Delta
t\rightarrow 0$), when a photon was emitted from the box. On the
other hand, the energy of the emitted photon can be obtained by
measuring the difference in the box's mass. This seemed to result in
$\Delta E\Delta t\rightarrow 0$, and violate the uncertainty
relation for energy and time.

However, as  pointed out by Bohr\cite{photon-box1,photon-box2},
Einstein neglected the time-dilation effect, and then his deduction
was incorrect. Following from general relativity, the time-dilation
is attributed to the difference in the gravitational potential. Two
clocks  tick at different rates if they are at different heights.
For the clock in the box, the time uncertainty due to the vertical
position uncertainty $\Delta x$ is given
by\cite{photon-box1,photon-box2}
\begin{eqnarray}\label{dilation1}
\Delta t=\frac{g\Delta x}{c^2}t,
\end{eqnarray}
where $t$ denotes a period of weighting the photon. Bohr argue that
 the accuracy of the energy of the photon is restricted as
 \begin{eqnarray}\label{accuracy0}
 \Delta E\geq\frac{c^2\hbar}{gt\Delta x}.
 \end{eqnarray}
  When Eq. (\ref{dilation1}) is considered, the inequality (\ref{accuracy0}) gives rise to the uncertainty relation $\Delta E\Delta
t\geq\hbar$, and the consistency of quantum theory is still
 maintained. Obviously, gravity plays an important role in the
 Bohr's argument.

 In the following,  GUP will be considered along the line of Bohr's
argument, and its significance to gravitation will be  analyzed in
the gedaken experiment for weighting the photon. We first read the
original position of the pointer on the box before the shutter
opens. After the photon is released, the pointer moves higher than
its original position. In order to lower the pointer to its original
position, we hang some little weights on the box.  The pointer
returns to its original position after  a period $t$.   The photon's
weight $g\Delta m$ equals the total weight
 that hangs on the box. Obviously, the accuracy of weighting the photon is determined by the minimum of the added weight. The
 measurement
 becomes meaningless, if the added weight is too small to be
 observable. The weight $g\Delta m$ should be restricted by quantum
 theory. Let $\Delta x$ denote the accuracy of measuring the position of the
 pointer(or of the clock), the minimum of the  momentum uncertainty is given by
\begin{eqnarray}
\Delta p_{min}=\frac{z\hbar}{\Delta x},
\end{eqnarray}
where  (\ref{uncertainty2}) has been considered. Over a period $t$,
the smallest weight is $\Delta p_{min}/t$, which is the quantum
limit of weighting the photon. Thus we obtain
\begin{eqnarray}
\frac{z\hbar}{t\Delta x}=\frac{\Delta p_{min}}{t}\leq g\Delta
m,\nonumber
\end{eqnarray}
or
\begin{eqnarray}\label{zhbar1}
z\hbar=\Delta x\Delta p_{min}\leq \Delta m(g\Delta x)t.
\end{eqnarray}
Substituting (\ref{dilation1}) into (\ref{zhbar1}), the latter
becomes
\begin{eqnarray}\label{uncertainty3}
z\hbar\leq c^2\Delta m\Delta t=\Delta E\Delta t.
\end{eqnarray}
Such a modified uncertainty relation means the shortest interval of
the time. It can be explained as follows. Since GUP is required to
ensure the shortest observable length, we have (\ref{uncertainty2}),
and then $z\hbar\geq\ell_\textsc{p}/\Delta p$. The time uncertainty
is restricted by $\Delta t\geq z\hbar/\Delta
E\geq\ell_\textsc{p}(\Delta p/\Delta E)\approx
\ell_\textsc{p}(dp/dE)$. $dE/dp$ is the speed of the box, so we
obtain $\Delta
t\geq\ell_\textsc{p}/v\geq\ell_\textsc{p}/c=t_\textsc{p}=\sqrt{G\hbar/c^5}$.

Now we turn our attention to the inequality (\ref{zhbar1}), from
which the  accuracy of the energy of the emitted photon is
restricted as
\begin{eqnarray}\label{accuracy}
\Delta E\geq\frac{z c^2\hbar}{gt\Delta
x}=\frac{c^2\hbar}{g^{\prime}t\Delta x},
\end{eqnarray}
Comparing it with (\ref{accuracy0}), the difference is only that $g$
is replaced by $g^{\prime}=g/z$. Let us define $p^{\prime}=\hbar k$,
which denotes the canonical momentum and satisfies Heisenberg's
commutation relation as follows
\begin{eqnarray}
[\hat{x},\hat{p}^{\prime}]=i\hbar.
\end{eqnarray}
 The corresponding uncertainty $\Delta p^{\prime}$ can be expressed as
\begin{eqnarray}\label{uncertainty4}
\Delta p^{\prime}=\hbar\Delta k\approx\hbar\frac{dk}{dp}\Delta
p=\Delta p/z,
\end{eqnarray}
where (\ref{mdb3}) has been considered. On the other hand, the
inequality (\ref{zhbar1}) can be rewritten as
\begin{eqnarray}\label{zhbar3}
\hbar=\Delta x\Delta p_{min}/z\leq \Delta m(g\Delta x)t/z.
\end{eqnarray}
Considering (\ref{uncertainty4}) and (\ref{zhbar1}), we have
\begin{eqnarray}\label{zhbar4}
\hbar=\Delta x\Delta p^{\prime}_{min}&=&\Delta x\Delta
p_{min}/z\nonumber\\
 &\leq& \Delta m(g\Delta x)t/z=\Delta
m(g^{\prime}\Delta x)t.
\end{eqnarray}
We find that $g^{\prime}$ is reproduced in the inequality
(\ref{zhbar4}), accompany with the return of  Heisenberg principle.
The means that $g^{\prime}$ should be understood in the context of
usual quantum theory. According to the formula (\ref{dilation1}),
when $g\rightarrow g^{\prime}$, the time uncertainty becomes
\begin{eqnarray}\label{dilation2}
\Delta t^{\prime}=\frac{g^{\prime}\Delta x}{c^2}t=\Delta t/z.
\end{eqnarray}
Substituting it into (\ref{zhbar4}), we obtain
\begin{eqnarray}\label{zhbar5}
\hbar=\Delta x\Delta p^{\prime}_{min}\leq \Delta E\Delta t^{\prime},
\end{eqnarray}
as required by usual quantum theory. The consistency of  theory is
maintained by the transformation: $g\rightarrow g^{\prime}$, when
the GUP's significance is explained in the context of  usual quantum
mechanics.

In summary, we suggest two pictures for understanding the COW phase
shift and the gedanken experiment of weighting the photon. One
picture is that GUP is considered directly in a classical
gravitational field. In another picture,  the usual quantum theory
is retained by introducing the effective gravitational field
strength $g^{\prime}$. Two pictures are equivalent, since an
observer cannot distinguish the effect of GUP from the effective
field strength. This equivalence inspires an effective Newton
constant $G^{\prime}$, since the effective field strength can be
expressed as
\begin{eqnarray}
g^{\prime}=g/z=\frac{GM}{zR^2}=\frac{G^{\prime}M}{R^2}, \nonumber
\end{eqnarray}
where $G^{\prime}=G/z$ is just the same as the suggestion in
Ref.\cite{lx2}.

In view of the above analysis,  we introduce  two working
hypotheses: (i) the matters obey Heisenberg's uncertainty principle;
(ii) quantum spacetimes are characterized by the GUP-inspired
effective Newton constant. They are the basis for the following
discussions. When $G$ is replaced by $G^{\prime}$, we
 obtain a modified Schwarzschild metric as follows
 \begin{eqnarray}\label{metric1}
 ds^2&=&-\left(1-\frac{2G^{\prime}M}{c^2r}\right)c^2dt^2+\left(1-\frac{2G^{\prime}M}{c^2r}\right)^{-1}dr^2+r^2d\Omega,\\
G^{\prime}&=&G/z.\nonumber
 \end{eqnarray}
 This metric describes a family of spacetimes that depend on
different scales of momentum. In the following section, we will
propose a characteristic momentum to incorporate quantum effect with
geometric character of  gravity.

\section{Gravitational tidal force and the characteristic momentum}
Now we consider a black hole described by (\ref{metric1}), with
$z=1+\alpha \ell_\textsc{p}^2 p^2/\hbar^2$. Let $T$ denote the black
hole temperature, and  the characteristic momentum is identified
with $k_BT/c$\cite{ling1,ling2}, the metric (\ref{metric1}) becomes
\begin{eqnarray}\label{metric2}
 ds^2&=&-\left(1-\frac{2G\widetilde{M}}{c^2r}\right)c^2dt^2+\left(1-\frac{2G\widetilde{M}}{c^2r}\right)^{-1}dr^2+r^2d\Omega,\\
 \widetilde{M}&=&\frac{M}{1+\alpha\ell_\textsc{p}^2k_B^2 T^2/c^2\hbar^2}.\nonumber
 \end{eqnarray}
The horizon is located by $r_{T}=2G\widetilde{M}/c^2$.  The
temperature, proportional to the surface gravity, is determined by
\begin{eqnarray}
T=\frac{m_\textsc{p}^2c^2}{8\pi k_B
\widetilde{M}}=\frac{m_\textsc{p}^2c^2}{8\pi k_B
M}\left(1+\frac{\alpha \ell_\textsc{p}^2k_B^2
T^2}{c^2\hbar^2}\right),
\end{eqnarray}
where $m_\textsc{p}=\sqrt{\hbar c/G}$ is the Planck mass.  So we
obtain
\begin{eqnarray}\label{temperature1}
T=\frac{4\pi M-\sqrt{(4\pi M)^2-\alpha m_\textsc{p}^2}}{\alpha
k_B/c^2},
\end{eqnarray}
which returns to the usual formula $T_H=m_\textsc{p}^2c^2/(8\pi k_B
M)$, as $\alpha\rightarrow 0$. Except an inessential factor, the
modified temperature (\ref{temperature1}) is consistent with the
previous work in the
literarure\cite{pschen1,vagenas2,lx1,ling1,ling2}. Certainly, those
old problems have yet not been solved. The expression
(\ref{temperature1}) gives rise to
 the maximum temperature  when  the mass approaches the minimal value $\sqrt{\alpha}m_\textsc{p}/{4\pi}$. Furthermore, the metric (\ref{metric2})
indicates that there is still a singularity at $r=0$. However, the
minimal mass means a lower bound on the size of black hole, $r_T\geq
\sqrt{\alpha}\ell_\textsc{p}/{4\pi}$. It is of order of the shortest
observable distance derived from GUP. This make us believe that
quantum spacetime is still characterized by the effective Newton
constant $G^{\prime}=G/z$, if an appropriate characteristic scale is
taken into account. The shortage of the  metric (\ref{metric2}) may
be attributed to the fact that the black hole temperature is not an
universal scale  of meaning, since it can't  describe an ordinary
star. Moreover, the black hole temperature is position independent,
and doesn't reflect  the difference between strong gravitational
field and weak field. It is necessary to reinvest the momentum scale
with new physical meaning, if we expect for something beyond the
previous efforts.  As an observable quantity, $p^2\geq\Delta p^2$.
In view of the limitation of the black hole temperature, we require
the quantum fluctuation $\Delta p$ to satisfy some reasonable
expectations.

Firstly, $\Delta p$ should be associated with the gravity, and
should increase with the strength of the gravitational field, i.e.
$\Delta p\sim r^{-s}$, $s>0$.

Secondly, $\Delta p$  should play a crucial role that improves the
spacetime singularity, and make the black hole regular. As argued in
Refs.\cite{reuter,hayward,lx2},  the asymptotic behavior of the
regular potential must satisfy $\phi\rightarrow r^{2+\delta}$, as
$r\rightarrow 0$. This demands $\Delta p\sim r^{-{3/2-\delta}}$, and
$\delta\geq 0$.

Thirdly, the minimal value of $\Delta p$ is intrinsic,  and reflect
the universal property of the gravitational fields of black holes
and ordinary stars. According to general relativity, the
gravitational field is regarded as the curved spacetime
characterized by Riemann tensor $R_{\rho\lambda\mu\nu}$. This
suggests $\Delta p$ be associated with those quantities constructed
by Riemann tensor. Such a characteristic momentum may be estimated
by combining the gravitational tidal force with quantum theory,
since the tidal force is associated with the curvature of
spacetime\cite{mtw, liang}.

Let us consider a pair of virtual particles with energy $\Delta E$.
 When the virtual
particles are separated by a distance $\Delta x$, according to the
geodesic deviation equation, the tidal force reads\cite{mtw,
liang,nalikar}
\begin{eqnarray}
F=\frac{2GM}{r^3}\left(\frac{\Delta E}{c^2}\right)\Delta x.
\end{eqnarray}
Let $\Delta t$ denote the life-time of the virtual particles,  the
momentum uncertainty due to the tidal force is given by
\begin{eqnarray}\label{tidal1}
\Delta p=F\Delta t=\frac{2GM}{r^3}\left(\frac{\Delta
E}{c^2}\right)\Delta t\Delta x.
\end{eqnarray}
The observability requires $\Delta p\Delta x\geq\hbar,~\Delta
E\Delta t\geq\hbar$, when the virtual particles are subject to the
tidal force and become real. Following from (\ref{tidal1}), we
obtain
\begin{eqnarray}\label{tidal2}
(\Delta p)^2&\geq&\frac{\hbar\Delta p}{\Delta x}=\frac{\hbar F}{\Delta x}\Delta t\nonumber\\
&=&\frac{2\hbar}{c^2}\left(\frac{GM}{r^3}\right)\Delta E\Delta
t\geq\frac{2\hbar^2}{c^2}\left(\frac{GM}{r^3}\right).
\end{eqnarray}
The right hand side of the  inequality suggests a characteristic
momentum, $\Delta p_{m}\sim\sqrt{GM\hbar^2/c^2r^3}$. We find that
$\Delta p_{m}\rightarrow 0$ as $G\rightarrow 0$, which is associated
with  a free particle traveling in the flat spacetime. This
characteristic scale can be understood as the minimal momentum of
those particles produced from the quantum fields in the curved
spacetime.

We can also analyze a  real particle which is detected by a photon
with energy $\Delta E$. Let $\Delta x$ denote the uncertainty in the
position of the particle, the momentum uncertainty of the particle
is given by
\begin{eqnarray}\label{tidal3}
\Delta \tilde{p}\geq\frac{\hbar}{\Delta x}+\frac{2GMm}{r^3}\Delta
x\Delta t,
\end{eqnarray}
where $m$ is the mass of the particle, and  $\Delta
t\geq\hbar/\Delta E$ is the characteristic time in the process of
the photon-particle collision. On the right hand side of the
inequality (\ref{tidal3}), the first and the second terms belong
different stories respectively. The second term is attributed to the
tidal effect of gravity, and it vanishes in a flat spacetime.
Following from (\ref{tidal3}), we obtain
\begin{eqnarray}\label{tidal4}
\Delta \tilde{p}\geq 2\sqrt{\frac{2G M\hbar m\Delta t}{r^3}}\geq
2\sqrt{\frac{2GM\hbar^2}{r^3}\left(\frac{m}{\Delta E}\right)},
\end{eqnarray}
where the time-energy uncertainty relation is considered. In order
to avoid the production of new particles, the energy should be
restricted as $\Delta E< mc^2$, otherwise it becomes meaningless to
measure the position of the particle\cite{lifshitz}. So we obtain
\begin{eqnarray}\label{tidal5}
\Delta \tilde{p}> 2\sqrt{\frac{2GM\hbar^2}{c^2r^3}}.
\end{eqnarray}
The difference between (\ref{tidal2}) and (\ref{tidal5}) is only a
 constant coefficient. This inessential difference is caused by a rough  estimate of the amplitude of $\Delta E$. It will vanish,
 if we take a different estimate, such as $\Delta E<mc^2/4$.

In the above discussion, we don't discriminate the black hole from
an ordinary star, so the characteristic momentum appeared in
(\ref{tidal2}) is suitable for both of the two spacetimes. A similar
scale has also been suggested by a different way\cite{reuter}, but
its physical meaning is different from our understanding.  As one of
two scales suggested in Ref.\cite{reuter}, it is identified with the
inverse of the proper time of an observer falling into the
Schwarzschild black hole. Obviously, it is different from that scale
of an ordinary star.

Identifying the characteristic scale with the right hand side of
(\ref{tidal2}), we obtain an effective Newton constant as follows
\begin{eqnarray}\label{rung2}
G^{\prime}=\frac{G}{1+2\alpha\ell_\textsc{p}^2\Delta
p_m^2/\hbar^2}=\frac{G}{1+2\alpha\ell_\textsc{p}^2GM/r^3c^2}.
\end{eqnarray}
Different from Ref.\cite{reuter},  the scale $\Delta p\sim 1/r$ is
no longer considered in the following discussions. This is because
it is motivated by Heisenberg principle, while Heisenberg principle
has been incorporated with the tidal effect in the inequalities
(\ref{tidal2}) and (\ref{tidal3}). We will explore a black hole
characterized by the effective Newton constant as presented by
(\ref{rung2}).

\section{Quantum tunneling from the regular black hole}
In this section, we take the Planck units, $G=\hbar=c=k_B=1$.
Substituting (\ref{rung2}) into (\ref{metric1}), we obtain a
modified Schwarzschild black hole as follows
\begin{eqnarray}\label{metric3}
ds^2&=&-\left(1+2\Phi\right)dt^2+\left(1+2\Phi\right)^{-1}dr^2+r^2d\Omega,\\
\Phi&=&-\frac{Mr^2}{r^3+2\alpha M}.\nonumber
\end{eqnarray}
It is just the minimal model of the regular black hole suggested in
Ref.\cite{hayward}. Let $\rho$ denote the radius of this black hole
and satisfy $g^{11}(\rho)=0$, we have
\begin{eqnarray}\label{horizon1}
1-\frac{2M\rho^2}{\rho^3+2\alpha M}=0.
\end{eqnarray}
The horizon is located by
\begin{eqnarray}
\rho=\frac{2M}{3}+\frac{4M}{3}\cos\left[\frac{1}{3}\arccos\left(1-\frac{27\alpha}{8M^2}\right)\right],
\end{eqnarray}
provided $M\geq M_c=27\alpha/16$. When $M\leq M_c$, the metric
(\ref{metric3}) doesn't describe a black hole, since  the equation
(\ref{horizon1}) has no positive solution and then the horizon is
absent in this spacetime\cite{hayward}. The critical mass is the
lower bond on the mass of an object that forms a black hole.
Corresponding to this critical mass, there is a minimal radius of
the black hole, $\rho_{min}=4M_c/3=\sqrt{3\alpha}$. In the
following, we will investigate the quantum tunneling of this regular
black hole, and focus on the question of whether the tunneling
probability is regularized by the minimal length.

In the tunneling picture of black hole
radiation\cite{parik,parikh2}, the tunneling probability is
determined by the imaginary part of the action for a particle which
tunnels through the horizon along a classically forbidden
trajectory. At the zeroth order WKB approximation, the tunneling
probability is suppressed by the change in the Bekenstein-Hawking
entropy. This is consistent with the unitarity of quantum theory.
When the second order correction is considered, the tunneling
probability is given by\cite{zjy1}
\begin{eqnarray}\label{tunnel1}
\Gamma\sim\frac{\rho_i^2}{\rho_f^2}\exp[-2\textbf{Im}(S_0-S_2)],
\end{eqnarray}
where $\rho_i$ denotes the initial radius of black hole in the
tunneling process, and $\rho_f$ the final radius.   $S_0-S_2$ is the
action for a particle crossing the horizon from $\rho_i$ to
$\rho_f$. Concretely speaking, $S_0$ is associated with the zeroth
order term of WKB wave function, and $S_2$ is related to the second
order correction. The first order term $S_1$ doesn't appear in  the
imaginary part of the action, since it is real. In order to evaluate
the emission rate of the regular black hole, we first introduce the
Painleve type coordinate\cite{parik, vagenas4}
\begin{eqnarray}
\tilde{t}=t+\int\frac{\sqrt{-2\Phi}}{1+2\Phi}dr.\nonumber
\end{eqnarray}
The metric (\ref{metric3}) is therefore rewritten as
\begin{eqnarray}
ds^2=-(1+2\Phi)d\tilde{t}^2+2\sqrt{-2\Phi}d\tilde{t}dr+dr^2+r^2d\Omega.
\end{eqnarray}
It is appropriate for describing the particle which tunnels through
the horizon, since  the coordinate singularity has been removed.
Setting $ds^2=0=d\Omega$, we obtain the equation of the radial null
geodesics as follows
\begin{eqnarray}\label{geodesics1}
\dot{r}=\frac{dr}{d\tilde{t}}=1-\sqrt{-2\Phi},
\end{eqnarray}
where the ingoing geodesics is neglected. When a particle is emitted
from the black hole and the energy conservation is considered,  a
shell with energy $\omega^{\prime}$ travals in a spacetime of mass
$M^{\prime}=M-\omega^{\prime}$. So we have
\begin{eqnarray}\label{tildephi}
1+2\Phi&=&\frac{r^3-2M^{\prime}r^2+2\alpha M^{\prime}}{r^3+2\alpha M^{\prime}}\nonumber\\
&=&-\frac{r^2-\alpha}{r^3+2\alpha
M^{\prime}}\left(2M^{\prime}-\frac{r^3}{r^2-\alpha}\right).
\end{eqnarray}
The zero order action is given by
\begin{eqnarray}\label{action0}
S_0=\int_{\rho_i}^{\rho_f}{S_0^{\prime}}dr=\int_{\rho_i}^{\rho_f}p_rdr,
\end{eqnarray}
where
\begin{eqnarray}\label{pr1}
S_0^{\prime}=p_r=\int\frac{dM^{\prime}}{\dot{r}}=\int_{M}^{M-\omega}\frac{dM^{\prime}}{1-\sqrt{-2\Phi}}.
\end{eqnarray}
Substituting (\ref{tildephi}) into (\ref{pr1}), we obtain
\begin{eqnarray}
p_r&=&\int_{M}^{M-\omega}\frac{1+\sqrt{-2\Phi}}{1+2\Phi}dM^{\prime}\nonumber\\
&=&-\int_{M}^{M-\omega}\frac{(1+\sqrt{-2\Phi})(r^3+2\alpha
M^{\prime})}{(r^2-\alpha)[2M^{\prime}-r^3/(r^2-\alpha)]}dM^{\prime}.
\end{eqnarray}
There exists a singularity at $2M^{\prime}=r^3/(r^2-\alpha)$. In
order for the positive frequency modes to decay with
time\cite{parik}, we deform the contour into the lower half
$\omega^{\prime}$ plane, or into the upper half $M^{\prime}$ plane.
By residue theorem, we obtain
\begin{eqnarray}
p_r&=&\left(\frac{-i\pi}{2}\right)~\frac{2\times[r^3+\alpha r^3/(r^2-\alpha)]}{r^2-\alpha}\nonumber\\
&=&(-i\pi)\frac{r^5}{(r^2-\alpha)^2}.
\end{eqnarray}
Substituting it into (\ref{action0}), we get the imaginary part of
the action as follows
\begin{eqnarray}\label{ims0}
\textbf{Im}S_0=-\frac{\pi}{2}\left[r^2+2\alpha\ln(r^2-\alpha)-\frac{\alpha^2}{r^2-\alpha}\right]\Bigg{|}_{\rho_i}^{\rho_f}.
\end{eqnarray}
Following the procedure of WKB method applied in Ref.\cite{zjy1}, we
can also evaluate the higher order terms of the action. The first
order term is determined by the following equation
\begin{eqnarray}
S_1^{\prime}=-\frac{S_0^{\prime\prime}}{2S_0^{\prime}}=-\frac{r^2-5\alpha}{2r(r^2-\alpha)}.\nonumber
\end{eqnarray}
So we have
\begin{eqnarray}
S_1^{\prime\prime}=\frac{r^4-14\alpha
r^2+5\alpha^2}{2r^2(r^2-\alpha)^2},\nonumber
\end{eqnarray}
and then
\begin{eqnarray}
S_2^{\prime}=-\frac{S_1^{\prime
2}+S_1^{\prime\prime}}{2S_0^{\prime}}=(-i)\times\frac{3r^4-38\alpha
r^2+35\alpha^2}{8\pi r^7}.
\end{eqnarray}
The imaginary part of the second order term  is given by
\begin{eqnarray}\label{ims2}
\textbf{Im}S_2&=&\textbf{Im}\int_{\rho_i}^{\rho_f}S_2^{\prime}dr\nonumber\\
&=&\frac{1}{96\pi}\left(\frac{18}{r^2}-\frac{114\alpha}{r^4}+\frac{70\alpha^2}{r^6}\right)\Bigg{|}_{\rho_i}^{\rho_f}.
\end{eqnarray}
Substituting (\ref{ims0}) and (\ref{ims2}) into (\ref{tunnel1}), and
considering  $\rho_i^2/\rho_f^2=\exp\ln (\rho_i^2/\rho_f^2)$, we
obtain the tunneling probability accurate to the second order
correction, $\Gamma\sim e^{\Delta S}$, where
\begin{eqnarray}\label{deltas}
 \Delta S=\left[\pi r^2-\ln r^2
 +\frac{3}{8\pi
r^2}+2\alpha\pi\ln(r^2-\alpha)-\frac{\alpha^2\pi}{r^2-\alpha}-\frac{19\alpha}{8\pi
r^4}+\frac{35\alpha^2}{24\pi r^6}\right]\Bigg{|}_{\rho_i}^{\rho_f}.
\end{eqnarray}
In the consideration of the unitarity of quantum theory, $\Delta S$
should be understood as the change in the entropy of the regular
black hole. The entropy, including the first and the second order
corrections, reads
\begin{eqnarray}\label{regulars}
S=\pi \rho^2-\ln \rho^2
 +\frac{3}{8\pi
\rho^2}+2\alpha\pi\ln(\rho^2-\alpha)-\frac{\alpha^2\pi}{\rho^2-\alpha}-\frac{19\alpha}{8\pi
\rho^4}+\frac{35\alpha^2}{24\pi \rho^6},
\end{eqnarray}
where $\rho$ is the radius of the black hole. The first three terms
are similar to the expression for the Schwarzschild black
hole\cite{zjy1}, while the last four terms are new. New corrections
are relevant to the parameter $\alpha$, and denote the difference
between the regular black hole and the Schwarzschild black hole.

Let us make some remarks on the  expressions (\ref{deltas}) and
(\ref{regulars}). Let us consider the thermodynamical entropy of the
regular black hole. As the inverse period of the imaginary time of
the regular spacetime (\ref{metric3}), the black hole temperature is
given by
\begin{eqnarray}
T&=&\frac{1}{2\pi}\left(\frac{d\Phi}{dr}\right)_{r=\rho}\nonumber\\
&=&\frac{1}{8\pi
M}-\frac{\alpha}{2\pi\rho^3}=\frac{\rho^2-3\alpha}{4\pi\rho^3},
\end{eqnarray}
where  (\ref{horizon1}) has been considered. The thermodynamical
entropy is defined as
\begin{eqnarray}
S^{(0)}&=&\int\frac{dM}{T}=\int T^{-1}\left(\frac{dM}{d\rho}\right)d\rho\nonumber\\
&=&2\pi\int\frac{\rho^5d\rho}{(\rho^2-\alpha)^2}\nonumber\\
&=&\pi\left[\rho^2+2\alpha\ln(\rho^2-\alpha)-\frac{\alpha^2}{\rho^2-\alpha}\right],
\end{eqnarray}
which is different from (\ref{deltas}). However,  it is consistent
with the entropy derived from the zero order action of WKB method,
as shown by (\ref{ims0}). This is similar to the Schwarzschild black
hole: the zero order action for the tunneling particle is related to
the change in the Bekenstein-Hawking entropy\cite{parik,zjy1}.

For the Schwarzschild black hole, the emission rate accurate to the
second order approximation,  is determined by the first three terms
in (\ref{deltas}). Since  classical general relativity doesn't
restrict the size of black hole,  $\Delta S$ and  $\Gamma$ become
divergent as $\rho_f\rightarrow 0$. However, the tunneling
probability of the regular black hole is finite, because it is
regularized by the minimal radius of the horizon. This conclusion is
nontrivial, in view of the subtle relation between the entropy
expression (\ref{regulars}) and the minimal radius  $\rho_{min}$. If
$\rho_{min}$ is allowed to be less than $\sqrt{\alpha}$,  the
entropy would be ill defined for the black hole of radius
$\rho=\sqrt{\alpha}$, because of the divergence of the fourth and
the fifth terms in (\ref{regulars}). It is gratifying that
$\rho_{min}=\sqrt{3\alpha}>\sqrt{\alpha}$, and those dangerous terms
such as $(\rho^2-3\alpha)^{-1}$ don't appear in the entropy
expression.

According to the third law of thermodynamics, the entropy vanishes
when  a system of matter is in the ground state and its temperature
approaches zero. For a given excited state, the probability of the
transition to the ground state should be minimal, because it is
greatly suppressed by the change in the entropy. The regular black
hole has similar property,  if the parameter $\alpha$ is not too
small. This is because the entropy expression (\ref{regulars}) is a
monotonic increasing function of the horizon area.\footnote{It can
be shown by numerical method that $dS/d{\rho}
>0$, when the parameter satisfies
 $\alpha>0.024.$}  For an initial black hole with radius $\rho_i$, the minimal
 value of the tunneling probability  is given by
\begin{eqnarray}\label{mgamma}
\Gamma\sim \exp\left[-\pi \rho_i^2+\ln \rho_i^2
 -\frac{3}{8\pi
\rho_i^2}-2\alpha\pi\ln(\rho_i^2-\alpha)+\frac{\alpha^2\pi}{\rho_i^2-\alpha}+\frac{19\alpha}{8\pi
\rho_i^4}-\frac{35\alpha^2}{24\pi \rho_i^6}\right],
\end{eqnarray}
  which points to the black hole remnant with final radius $\rho_f=\sqrt{3\alpha}$.

In the expression  (\ref{regulars}), the fourth and the fifth terms
is a part of the thermodynamical entropy of the regular black hole.
It is interesting and confusing that they tend to cancel out the
similar corrections to the entropy of the Schwarzschild black hole,
such as the second and the third terms in (\ref{regulars}). This
fact might indicate a subtle correlation between quantum spacetimes
and the quantum matters, but we don't know how to explain it. We
also notice that the regular black hole gives rise to the higher
order corrections to the entropy, such as the last two terms in
(\ref{regulars}). We predict that the similar and opposite
corrections might appears in the tunneling probability of a
Schwarzschild black hole, when the fourth order WKB approximation is
considered.
\section{Summary and outlook}
This work involves two parts. The first part is devoted to the
question of what is the significance of the GUP for the quantum
spacetime. The answer may point to the a scale-dependent Newton
constant, which is motivated by analyzing the role that the GUP
plays in the COW phase shift and Einstein-Bohr's Gedanken for
weighting photon. It is consistent with our previous suggestion in
Ref.\cite{lx2}. The minimal model of the regular black hole can be
reproduced by considering the simplest  GUP and a momentum scale
associated with the tidal force. The second part is to calculate the
tunneling probability accurate to the second order WKB
approximation. The tunneling probability is regular, because the
black hole has a nonzero minimal radius. Not only this, the
tunneling probability of an initial black hole is minimized by the
black hole remnant, if the parameter $\alpha$ is of order of the
unity. In other words, the tunneling probability is minimal, if the
final state of the black hole is a remnant.

Let us consider the matter source of the regular black hole. In this
paper, the quantum spacetime is understood by connecting the GUP
with the running of Newton constant. It reflects the  quantum
gravitational effects on the classical spacetime. According to the
general theory of relativity, the effective stress-energy tensor can
be derived from Einstein's field equation, when the regular black
hole is regarded as an input. The quantum gravitational effects are
simulated by a matter fluid described by the effective stress-energy
tensor\cite{reuter,hayward}. We hope that this matter fluid can be
reproduced from the GUP dominated vacuum fluctuations. This problem
will be investigated in the future.

 Besides constructing the above regular black hole, we also explore a theory of modified
 gravity. This alternative theory is based on a generalization of the effective Newton constant, and it may be characterized by a
modified Hilbert-Einstein action as follows
\begin{eqnarray}\label{actiong1}
I^{\prime}=\int\frac{R-2\Lambda}{16\pi G^{\prime}}\sqrt{-g}d^4{x},
\end{eqnarray}
where $G=c=1$, $G^{\prime}=z^{-1}(p)$, and $\Lambda$ is the
cosmological constant. In order for the Lagrangian to be an
invariant, the momentum scale $p$ is restricted to be a scalar. For
a Schwarzshild spacetime, $p^2\sim M/r^3$, as argued in the section
3. This suggests that the characteristic momentum should be
identified as $p\sim\sqrt{K}$, and then $z=z(K)$, where $K$ is the
square root of the Kretschmann scalar. Considering the simplest
GUP[as given by (\ref{commut0})], the gravitational action can be
expressed as
\begin{eqnarray}\label{actiong3}
I=\frac{1}{16\pi G}\int(1+\gamma K)(R-2\Lambda)\sqrt{-g}d^4{x},
\end{eqnarray}
where $\gamma$ is a parameter, which is not necessarily the same as
that in the  metric (\ref{metric3}). The action (\ref{actiong3})
belongs to a class of more general theories of modified
gravity\cite{carr,clifton, gcj}, and then the field equation is
given by
\begin{eqnarray}\label{field1}
(1+\gamma K)(G_{\mu\nu}+\Lambda g_{\mu\nu})+\gamma H_{\mu\nu}=8\pi
T_{\mu\nu},
\end{eqnarray}
where $G_{\mu\nu}$ is the Einstein tensor, and
\begin{eqnarray}
H_{\mu\nu}=\frac{R-2\Lambda}{K}R_{\rho\lambda\sigma\mu}R^{\rho\lambda\sigma}_{~~~\nu}
+(g_{\mu\nu}\nabla_{\sigma}\nabla^{\sigma}-\nabla_{\mu}\nabla_{\nu})K-2\nabla_\rho\nabla_{\sigma}\left[\frac{R-2\Lambda}{K}R^{\rho~~\sigma}_{({\mu\nu})~}\right].
\end{eqnarray}
 Direct calculation shows that the metric
(\ref{metric3}) is not a solution for the field equation
(\ref{field1}). It is not strange, since  the metric (\ref{metric3})
and the equation (\ref{field1}) are suggested along different lines
of argument, even though they are motivated by the effective Newton
constant. However,  the regular spacetime (\ref{metric3}) has a de
Sitter core near $r=0$, which satisfies the field equation
(\ref{field1}). This implies that the field equation (\ref{field1})
permits the existence of the  regular black holes.  We also take
notice of those terms associated with the cosmological constant
$\Lambda$ in  (\ref{field1}), i.e. $(1+\gamma K)\Lambda g_{\mu\nu}$.
Usually, the first term $\Lambda g_{\mu\nu}$ is  utilized to cancel
out the huge contribution from the vacuum energy  on the right hand
side of the field equation, where the bare $\Lambda$ must be of
order of unity. Thus the second term of $\gamma K\Lambda$ play the
role of the effective cosmological constant. It is interesting that
this term is of order of the observed value. The modified gravity
with the  square root of Kretschmann scalar seems to be ignored in
the literature. The field equation (\ref{field1}) and relevant
problems will be discussed in detail elsewhere. We hope that the
spacetime singularities and the cosmological constant problem can be
improved in this alternative theory.
 \section*{Acknowledgments}
 X. Li thanks X. J. Yang and Q. J. Cao for their helps and useful discussions. This work
is supported by Natural Science Foundation of Science Foundation of
China(grants No.11373020, No.11575195), and Natural Science
Foundation of Zhejiang Province of China(grant No.LY14A030001).


\begin{thebibliography}{99}
\bibitem{mtw}C.W.Misner, K.S.Thorne, J. A. Wheeler, {\it Gravitation},
W.H.FREEMAN and Company, 1970.
\bibitem{ph}S.W.Hawking and G.F.R. Ellis, {\it The Large Scale Structure
of Spacetime}, Combridge Universe Press, 1973.
\bibitem{hawk}S. W. Hawking,  Commun. Math. Phys. 43(1975), 199.
\bibitem{parik} M. K. Parikh and F. Wilczek, Phys. Rev. Lett.
85(2000), 5042.
\bibitem{parikh2}M. K. Parikh, Gen.Rel.Grav.36(2004), 2419. arXiv:hep-th/0405160.
\bibitem{vagenas4}E. C. Vagenas, Phys.Lett. B559(2003), 65. arXiv:hep-th/0209185.
\bibitem{vagenas1} M. Arzano, A. J. M. Medved and E. C. Vagenas,
JHEP, 0509(2005), 037.
\bibitem{zjy2} J. Y. Zhang and Z. Zhao, Phys.Lett. B638 (2006), 110.
arXiv:gr-qc/0512153.
\bibitem{zjy1}Jingyi Zhang, Phys. Lett. B668(2008), 353.
\bibitem{garay1}L. J. Garay, Int. J. Mod. Phys. A10(1995), 145, and
references therein.
\bibitem{pschen1}R. J. Adler, P. Chen and  D. I. Santiago,  Gen. Rel. Grav. 33 (2001), 2101.
\bibitem{vagenas2} A. J. M. Medved and E. C. Vagenas,
Phys. Rev. D70(2004), 124021.
\bibitem{lx1}Li Xiang and X. Q. Wen, JHEP, 10(2009),046.
\bibitem{bardeen} J. Bardeen, Proc. GR5, Tbilisi, USSR(1968).
\bibitem{reuter} A. Bonanno and M. Reuter, Phys. Rev.
D62(2000),043008.
\bibitem{nicolini} P. Nicolini, A. Smailagic and E. Spallucci, Phys.
Letts. B632(2006), 547.
\bibitem{hayward} S. A. Hayward, Phys. Rev. Lett. 96(2006), 031103.
\bibitem{myung} Y. S. Myung, Y-W. Kim and Y-J. Kim, Phys. Letts.
B656(2007), 221.
\bibitem{lx2} Li Xiang, Yi Ling and You-Gen Shen, Int. J. Mod.
D22(2013), 1342016.
\bibitem{culetu} H. Culetu, Int. J. Theor. Phys. 54(2015), 2855. arXiv:1408.3334[gr-qc].
\bibitem{vagenas3} L. Balart, E. C. Vagenas, Phys. Lett. B730(2014), 14.
arXiv:1401.2136[gr-qc].
\bibitem{ghosh} M. Amir, S. G. Ghosh, Phys. Rev. D 94(2016), 024054. arXiv:1603.06382[gr-qc].
\bibitem{chang1}S. Benczik, L. N. Chang, D. Minic and et al, Phys.Rev.D66(2002),
026003.
\bibitem{rama}S. K. Rama, Phys. Letts. B519(2001), 103.
\bibitem{hh}U. Harbach and S. Hossenfelder, Phys.Lett. B632(2006),
379.
\bibitem{kempf}A. Kempf, G. Mangano and R.B. Mann,   Phys.
Rev.  D52(1995), 1108.
\bibitem{ahlu}D.V. Ahluwalia, Phys. Letts. A275(2000), 31.
\bibitem{vagenas5} S. Deb, S. Das and E. C. Vagenas, Phys. Lett. B755(2016),
17. arXiv:1601.07893[gr-qc].
\bibitem{cdl}C. Cohen-Tannoudji, B. Diu and F. Laoe, {\it Quantum
Mechanics}, Hermann, 1977.
\bibitem{cow1} Colella, A. W. Overhauser and S. A. Werner,  Phys. Rev. Lett. 34(1975),
1472.
\bibitem{cow2}D. M. Greenberg and A. W. Overhauser, Rev. Mod.
Phys.51(1979), 43.
\bibitem{cow3} Guang Hong, {\it Elementary Concepts in Quantum Mechanics}(in chinese),
High Education Press, 1990.
\bibitem{saha}A. Saha, Phys. Rev. D. 89, 025010 (2014).
arXiv:1306.4202[hep-th].
\bibitem{photon-box1}{\it Niels Bohr, Selected Works}. J. Kalckar
Ed.Vol.7. Elsevier, Amsterdam, 1996.
\bibitem{photon-box2}Y. Aharonov and D. Rohrlich, {\it Quantum Paradoxes,
quantum theory for the perplexed}, WILEY-VCH Verlag GmbH $\&$ CO.
KGaA, 2005.
\bibitem{ling1} Y. Ling, B. Hu and X. Li, Phys. Rev. D73(2006),
087702.
\bibitem{ling2}Y. Ling, X. Li, and H. B. Zhang, Mod. Phys. Letts.
A22(2007), 2749.
\bibitem{liang}C. B. Liang and B. Zhou, {\it Elementary Differential Geometry and
General Relativity}(in chinese), Scinence Press, 2006.
\bibitem{nalikar} J. V. Narlikar and T. Padmanabhan, {\it Gravity, Gauge Theories and Quantum
Cosmology}, Reidel, Dordrecht, 1986.
\bibitem{lifshitz}E. M. Lifshitz, L.P. Pitaevskii and V. B.
Berestetskii, {\it Quantum Electrodynamics}, Reed Educational and
Professional Publishing, 1982.
\bibitem{carr} S. M. Carroll, et al., Phys. Rev. D71(2005), 063513.
axXiv:astro-ph/0410031.
\bibitem{clifton}T. Clifton and J. D. Barrow, Phys. Rev. D72(2005),
123003.  arXiv:gr-qc/0511076.
\bibitem{gcj} C.J. Gao, Phys. Rev. D86(2012), 103512.   arXiv:1208.2790[gr-qc].
\end{thebibliography}
\end{document}